# Force autocorrelation function in linear response theory and the origin of friction

**Janka Petravic**

**Complex Systems in Biology Group, Centre for Vascular Research, The University of New South Wales, Sydney NSW 2052, Australia**

## *ABSTRACT*

Vanishing of the equilibrium Green-Kubo fluctuation expression for the friction coefficient of a massive particle moving in a finite-volume liquid is usually interpreted as an unphysical consequence of the finite volume. Here I show that it is a physical consequence of the finite mass of the rest of the system, which allows it to be dragged by the moving particle. As a consequence, it is sufficient to have two infinite masses in the liquid for the friction coefficient to be finite. In addition, I give the physical interpretation of different friction coefficients for two infinite-mass particles moving in the liquid.

## 1. INTRODUCTION

A rigid body, moving through or past a liquid at constant velocity, experiences resistance to its motion due to collisions with liquid molecules. The moving body can be a macroscopic solid, like for example a wall of a pipe, a suspended colloidal particle or even a molecule of the same liquid. If the motion is slow, friction coefficient $\zeta$, defined as the ratio of the force needed to overcome liquid resistance and the velocity of the rigid body, is a constant independent of velocity. An important result of statistical physics relates friction coefficient $\zeta$ to fluctuations of the force **F** acting on the body in equilibrium [1-6],

$$\zeta(t) = \frac{1}{3k_B T}\int_0^t \langle \mathbf{F}(s)\cdot\mathbf{F}(0)\rangle dt\,,$$

**Eq. 1**

where $k_B$ is the Boltzmann constant, $T$ is temperature, and triangular brackets $\langle\ldots\rangle$ denote the ensemble average. Eq. 1 expresses the friction coefficient as the Green-Kubo-type integral of the force autocorrelation function. However, a major problem with this expression is that it vanishes in the long-time limit if the volume of the liquid in which the rigid body is suspended is finite. This holds if the solid particle is in finite-volume liquid as well as in periodic boundary conditions.

At first sight, we see the vanishing of the Green-Kubo friction coefficient as contrary to experience: we assume that there will be fluid resistance at all times if we move a solid sphere through a liquid, even if liquid volume is finite. Since the expression Eq. 1 has been originally derived using stochastic methods involving a number of assumptions and approximations [1], it has been generally assumed that the method breaks down when the volume is finite, or in other words when the infinite time limit is taken before the infinite volume limit [1,6,7] (as is necessarily done in all molecular dynamics simulations).

However, Eq. 1 can be alternatively obtained using linear response theory, which is a straightforward method for obtaining an exact limit of the ratio of force and velocity, involving no approximations. The time-dependence of the friction coefficient Eq. 1 calculated in equilibrium should then reflect the true behaviour of the friction coefficient in time, in agreement with the Onsager regression hypothesis that equilibrium fluctuations relax in the same way as flux would relax to a steady state in a low field. Is then our image of what should happen with friction in finite volume at fault? I show below that a careful analysis of the nonequilibrium experiment for which Eq. 1 represents the linear limit shows that the vanishing of the integral in Eq. 1 for $t \to \infty$ is not a spurious consequence of finite volume, but a true description of the outcome in a finite-volume system in the long-time limit.

This result makes it possible to interpret the physical meaning of the elements of friction tensor for two rigid bodies in a liquid [8] and the true meaning of the friction force in Eq. 1. In particular, it allows us to interpret the finite values of the Green-Kubo force integrals for two massive particles moving in the liquid, and find that long-term friction can only exist when two particles are moving through liquid *relative* to each other.

## 2. ONE PARTICLE MOVING IN FINITE-VOLUME LIQUID

In this Section, I first review the linear response formalism, and then apply the general formulae to the special case of one particle moving at a constant velocity through a liquid. A particle which moves at a constant velocity irrespective of collisions with liquid atoms/molecules necessarily has infinite mass – the case most often considered in literature [9]. Finally, I discuss the reciprocal problem: a finite-volume liquid of finite-mass molecules initially moving relative to a stationary particle of infinite mass.

### A. Linear response theory

Consider a system of $N$ particles ($i$=1,…,$N$) of mass $m$ obeying the equations [10]

$$\dot{\mathbf{r}}_i = \mathbf{p}_i / m + \mathbf{C}_i(\mathbf{r},\mathbf{p}) F_{ext}$$
$$\dot{\mathbf{p}}_i = \mathbf{F}_i + \mathbf{D}_i(\mathbf{r},\mathbf{p}) F_{ext},$$

**Eq. 2**

where $\mathbf{r}_i$, $\mathbf{p}_i$ are their positions and momenta respectively, and $\mathbf{F}_i$ are the forces of interaction experienced by each particle. $\mathbf{C}_i(\mathbf{r},\mathbf{p})$ and $\mathbf{D}_i(\mathbf{r},\mathbf{p})$ are vector functions that can in general depend on all positions and momenta, and $F_{ext}$ is the magnitude of external field. $F_{ext}$ =0 corresponds to equilibrium. We assume that the field is applied at $t$=0, so that $F_{ext}$=0 for $t$<0, and constant in time. The flux $J$ generated by Eq. 2 satisfies the equation

$$\dot{H}_0 = -VJF_{ext},$$

**Eq. 3**

where $H$ is the equilibrium Hamiltonian of the system Eq. 2 and $V$ is the volume, and has the form

$$J = \frac{1}{V}\sum_{i=1}^{N}[\mathbf{F}_i \cdot \mathbf{C}_i - \frac{\mathbf{p}_i}{m} \cdot \mathbf{D}_i].$$

**Eq. 4**

Let $B(\mathbf{r},\mathbf{p})$ be an arbitrary function of positions and momenta with the vanishing equilibrium ensemble average $\langle B(\mathbf{r},\mathbf{p})\rangle$. When the vectors $\mathbf{C}_i$ and $\mathbf{D}_i$ do not depend explicitly on time and satisfy the “adiabatic incompressibility of phase space”,

$$\sum_{i=1}^{N}[(\partial / \partial \mathbf{r}_i) \cdot \mathbf{C}_i + [(\partial / \partial \mathbf{p}_i) \cdot \mathbf{D}_i] = 0,$$

**Eq. 5**

one can evaluate the exact limit $\langle B(\mathbf{r},\mathbf{p})\rangle/F_{ext}$ at time $t$ as $F_{ext} \rightarrow 0$ from the equilibrium fluctuations [10],

$$\lim_{F_{ext}\rightarrow 0}\frac{\langle B(t)\rangle}{F_{ext}} = -\frac{V}{k_B T}\int_0^t \langle B(s)J(0)\rangle ds \,.$$

**Eq. 6**

The left hand side of Eq. 6 is the transport coefficient in the linear response regime associated with the quantity $B$ in the nonequilibrium process created by $F_{ext}$. The limit is exact – it contains no approximations. As long as we calculate the equilibrium correlation integral on the right hand side from the equations of motion Eq. 2 with $F_{ext}$ =0, the right hand side of Eq. 6 will describe the exact value of the ratio of the ensemble average of the quantity $B$ and the applied field $F_{ext}$ for sufficiently low values of $F_{ext}$. The value of the integral at any time $t$ would depend on the type of the flow, the choice of $B$ and the thermodynamic properties of the system (e.g. $N$, $V$, $T$). Linear response theory does not predict the value of the integral for any time $t$, but if the nonequilibrium process gives rise to a non-vanishing steady-state value of $\langle B\rangle$, then the right hand side will not vanish as $t\rightarrow\infty$. In contrast, if the change in $\langle B\rangle$ is only transient, then the right hand side will vanish at long times.

## B. Application to a particle suspended in a liquid

Let us now apply the linear response formalism to a case of a particle ($i$=1) of mass M suspended in the liquid of $N$-1 particles of mass $m$. The external field acts only on the particle 1 and is a constant velocity $v_0$ in $x$-direction added to its equilibrium velocity. The equations of motion with the external field become:

for particle one: for liquid atoms/molecules ($i$=2,…,$N$):

$$\dot{\mathbf{r}}_1 = \mathbf{p}_1 / M + \mathbf{e}_x v_0 \qquad \dot{\mathbf{r}}_i = \mathbf{p}_i / m$$
$$\dot{\mathbf{p}}_1 = \mathbf{F}_1 \qquad \dot{\mathbf{p}}_i = \mathbf{F}_i ,$$

**Eq. 7**

so that $F_{ext}= v_0$, $\mathbf{C}_1= \mathbf{e}_x$ (the unit vector in $x$-direction) and $\mathbf{C}_i(i=2,\ldots,N) = \mathbf{D}_i(i=1,\ldots,N) = 0$. According to Eq. 4, the dissipative flux $J$ is equal simply to the interaction force acting on particle 1 in $x$-direction, divided by volume,

$$J = F_{x1} / V .$$

**Eq. 8**

$F_{x1}$ is the force experienced by the particle 1, opposing its motion in $x$-direction at velocity $v_0$, i.e. the force of friction. Since the system Eq. 7 satisfies the adiabatic incompressibility condition Eq. 5, the linear response (slow $v_0$) friction coefficient is according to Eq. 6 equal to

$$\lim_{F_{ext}\to 0} \frac{\langle F_{x1}(t)\rangle}{v_0} = -\frac{1}{k_B T}\int_0^t \langle F_{x1}(s)F_{x1}(0)\rangle ds.$$

**Eq. 9**

Using stochastic methods [1] or the generalized Langevin equation approach [7] one can find that the condition for the integral not to vanish as its upper limit goes to infinity [7] is

$$\frac{M(N-1)m}{M+(N-1)m} \to \infty ,$$

**Eq. 10**

which amounts to $M\to\infty$ (infinite mass of particle 1) and $N\to\infty$ (thermodynamic limit) simultaneously. Since $M\to\infty$ is one of the necessary conditions for the right hand side of Eq. 9 not to vanish at long times, this is the case investigated in all the previous studies [1-9].

When the mass of the suspended particle 1 is infinite, then it is strictly immobile in equilibrium irrespective of the collisions with the liquid molecules (its velocity is zero but its momentum is not, so that total momentum of the whole system is still formally conserved). If the integral Eq. 9 is evaluated with particle one at rest and all the other particles moving according to Newton's equations, it should represent the friction coefficient for a particle moving at a constant velocity through a liquid. Sure enough, in finite volume (e.g. in periodic boundary conditions) the integral decays exponentially in time, implying that friction gradually disappears as the particle moves.

We can alternatively study the opposite problem - finite-mass liquid molecules initially moving relative to a stationary particle of infinite mass [7]. The equations of motion for this situation would be

for particle 1: for liquid atoms/molecules ($i$=2,…,$N$):

$$\dot{\mathbf{r}}_1 = \mathbf{p}_1 / M \qquad \dot{\mathbf{r}}_i = \mathbf{p}_i / m + \mathbf{e}_x v_0$$
$$\dot{\mathbf{p}}_1 = \mathbf{F}_1 \qquad \dot{\mathbf{p}}_i = \mathbf{F}_i ,$$

**Eq. 11**

with dissipative flux

$$J = \sum_{i=2}^{N} F_{xi} / V = -F_{x1} / V ,$$

**Eq. 12**

so that the force of friction acting on particle 1 would be

$$\lim_{F_{ext} \to 0} \frac{\langle F_{x1}(t) \rangle}{v_0} = \frac{1}{k_B T} \int_0^t \langle F_{x1}(s) F_{x1}(0) \rangle ds,$$

**Eq. 13**

i.e. of opposite sign from when the particle 1 is moving. Thus, vanishing of the integral Eq. 9 in the long time limit seems to also imply that friction force felt by a particle immersed in a liquid moving past it at constant velocity will disappear in time if the volume of liquid is finite.

These results appear counterintuitive, and yet they should be the real outcome of the nonequilibrium processes described by Eq. 7 and Eq. 11. Next, I use a combination of equilibrium and nonequilibrium molecular dynamics methods to show that friction really does disappear for a particle moving at constant speed through a finite-volume liquid.

## C. Simulation results

The system consisted of 256 Lennard-Jones (LJ) atoms in periodic boundary conditions, at the number density $\rho$=0.82 and temperature $T$=1.0 in the LJ system of reduced units. In equilibrium simulations atom 1 was immobile, while the rest had mass $m$ equal to unity and obeyed Newton's equations of motion with Nosé-Hoover thermostat. The equations of motion were solved using 5$^{th}$ order Gear integrator with the time step of 0.001. The values of force on atom 1 were recorded every 5 time-steps. The correlation function in Eq. 9 was calculated using the shifting register technique [11] with a time-window of 15000 records (or 75000 time-steps). Each recorded value was used as an initial point for the averaging of the correlation function. The result was averaged over all three Cartesian directions.

In the first set of nonequilibrium simulations, atom 1 moved at the constant velocity of 0.3 LJ reduced units in $x$-direction, while the rest of the system obeyed Newton's equations of motion Nosé-Hoover thermostat applied in $y$- and $z$-direction ($x$-direction of motion was not thermostatted in order not to disturb the resulting flow). Temperature stayed constant throughout each of nonequilibrium runs. 5000 nonequilibrium runs started from equilibrium configurations separated by $10^4$ time-steps. An additional 5000 configurations were obtained from the original initial equilibrium configurations by mapping $x_i \rightarrow -x_i$ and $p_{xi} \rightarrow -p_{xi}$ (in order to ensure that the starting ensemble averages of liquid velocity and force on atom 1 were zero). Force on atom 1 ($F_{x1}$) and flow velocity of the rest of the system $v_{liq}$ were averaged at every time-step over the resulting $10^4$ nonequilibrium runs.

Fig.1a illustrates the simulated nonequilibrium process. A system in periodic boundary conditions is essentially an infinite system consisting of identical building blocks. Specifically, our infinite mass in periodic boundary conditions represents an infinite-mass simple cubic lattice, that at $t$=0 starts to move with velocity $v_0$ through a liquid. It is easy to predict what is going to happen in this system in time. At first, the liquid surrounding the infinite-mass particle will be at rest, and it will have to push the liquid out of its way in order to move, resulting in the force of friction opposing its motion. As it moves, it will transfer momentum to the surrounding liquid, and drag the liquid atoms/molecules with it. After sufficiently long time has passed, liquid will be moving together with the infinite-mass lattice. There will be no relative motion of the lattice and the rest of the liquid, and consequently no friction force. In Fig.1b I compare the equilibrium friction coefficient (right hand side of Eq. 9; full line) and the ratio of the force on the infinite-mass particle and its velocity obtained in direct nonequilibrium simulation (dashed line). Force decays exponentially in both cases, and equilibrium and nonequilibrium results can hardly be distinguished. In Fig.1c I plot the ensemble average of the force on particle 1, as it moves through the liquid, against the average liquid velocity. By the time the force on the moving lattice decays to zero, the average velocity of the rest of the liquid has become equal to $v_0$, the constant velocity of the lattice. Since no relative motion exists between the infinite-mass particle and the liquid, there is no more friction.

In the reciprocal nonequilibrium situation, the finite-mass liquid atoms are assigned an additional velocity $v_0$ in $x$-direction, while the infinite-mass particle is at rest (Fig.1d), as described by Eq. 11 with $M\to\infty$. The force on the stationary particle divided by $v_0$ again follows the equilibrium result, but this time with the opposite sign (Fig.1e).

Fig.1f shows the decay of the force against the average liquid velocity $v_{liq} = \dot{x}_{liq}$, which also decays to zero. In Eq. 11, the momenta of the liquid atoms differ from their velocities divided by mass. It is the velocity of the liquid atoms that governs their trajectory (how much distance they cover in unit time), and therefore also their interactions. Interactions in turn change their momenta. Initially, velocities are preferably in $x$-direction, so that in collisions with the stationary particle there is more backscattering than forward scattering. The final outcome is that the average liquid velocity relaxes to zero. In this process, the liquid atoms have acquired the average negative momentum equal to $-v_0/m$. The stationary, infinite-mass particle has acquired an average finite positive momentum of $(N\text{-}1)v_0/m$, which does not cause it to move because of its infinite mass.

The presented analysis of equilibrium and nonequilibrium simulations clarifies why it is impossible to obtain an equilibrium expression for a friction coefficient as a long-time limit of the correlation integral Eq. 1, where the correlation function in the integrand on the right hand side is evaluated for a stationary particle in a finite volume. This expression represents the linear limit of the force on a particle moving at a constant unit velocity, i.e. on a particle of infinite mass. Such a particle will necessarily finally drag the rest of the liquid with it, so that the velocities of the moving particle and the surrounding liquid will match. Friction is the consequence of motion of the particle *relative* to liquid, and disappears as the relative velocity decays to zero.

## *3. TWO PARTICLES IN FINITE-VOLUME LIQUID*

Let us examine again the condition for the finiteness of the integral Eq. 1 in the limit $t\rightarrow\infty$, as presented in Eq. 10. This condition can be rewritten as

$$\frac{MM_{liq}}{M+M_{liq}} \to \infty ,$$

**Eq. 14**

where $M_{liq}$ is the total mass of the rest of the system. When $M\to\infty$ but $M_{liq}$ is finite, the one infinite mass can drag the whole rest of the system, and the friction coefficient Eq. 1 decays to zero. When the rest of the system also has infinite mass and is initially at rest (has zero average velocity), one particle moving at constant speed will not be able to drag it, so that there will always be relative motion between the moving particle and the liquid. $M_{liq}$ will be infinite in the thermodynamic limit $N\to\infty$, but this is not the only possibility. It is sufficient for the rest of the system to contain at least another one infinite-mass particle, even when the total number of atoms $N$ is finite.

## A. Both particles are moving together at the same speed

Let us consider a system consisting of two particles ($i$=1,2) of infinite mass $M$ and $N$-2 particles of finite mass $m$ ($i$=3,…,N). In periodic boundary conditions this corresponds to two simple cubic lattices of infinite mass displaced with respect to each other by $\mathbf{r}_2 - \mathbf{r}_1$ (see Fig.2). If at time $t$=0 both particles start to move with the same velocity $v_0$ in the $x$-direction, the equations of motion will be

for particles 1,2: for liquid atoms ($i$=3,…,$N$):

$$\dot{\mathbf{r}}_{1,2} = \mathbf{p}_{1,2}/M + \mathbf{e}_x v_0 \qquad \dot{\mathbf{r}}_i = \mathbf{p}_i/m$$
$$\dot{\mathbf{p}}_{1,2} = \mathbf{F}_{1,2} \qquad \dot{\mathbf{p}}_i = \mathbf{F}_i ,$$

**Eq. 15**

with dissipative flux

$$J = -\sum_{i=2}^{N} (F_{x1} + F_{x2})/V$$

**Eq. 16**

leading to linear response friction forces on particles 1, 2

$$\frac{\langle F_{x1,2}(t)\rangle}{v_0} = -\frac{1}{k_B T}\int_0^t \langle F_{x1,2}(s)[F_{x1}(0)+F_{x2}(0)]\rangle ds\,.$$

**Eq. 17**

The simultaneous motion of two infinite-mass particles through the liquid will result with the same outcome as the motion of one infinite-mass particle discussed previously. As the liquid eventually acquires the same velocity as the two particles, friction force on each of the particles will disappear, because only motion relative to the liquid generates friction. The simulation results for the decay of the equilibrium friction forces on 1 and 2, obtained in the same system as described in previous section (but now with two infinite-mass LJ atoms) is shown in Figure 2b.

## B. One particle is moving, the other is stationary

Let us now consider a nonequilibrium system depicted in Fig.3a, where particle 1 (black lattice) starts moving at velocity $v_0$ in the $x$-direction at $t$=0, while particle 2 (white lattice) remains stationary. Nonequilibrium equations of motion are

| for particle 1 | for particle 2 | for liquid atoms ($i$=3,…,$N$): |
|---|---|---|
| $\dot{\mathbf{r}}_1 = \mathbf{p}_1/M + \mathbf{e}_x v_0$ | $\dot{\mathbf{r}}_2 = \mathbf{p}_2/M$ | $\dot{\mathbf{r}}_i = \mathbf{p}_i/m$ |
| $\dot{\mathbf{p}}_1 = \mathbf{F}_1$ | $\dot{\mathbf{p}}_2 = \mathbf{F}_2$ | $\dot{\mathbf{p}}_i = \mathbf{F}_i,$ |

**Eq. 18**

with dissipative flux

$$J = -\sum_{i=2}^{N} F_{x1}/V$$

**Eq. 19**

and the corresponding friction coefficients,

$$\frac{\langle F_{x1,2}(t)\rangle}{v_0} = -\frac{1}{k_B T}\int_0^t \langle F_{x1,2}(s)F_{x1}(0)\rangle ds\,.$$

**Eq. 20**

Simulation results for the equilibrium friction coefficients from Eq. 20 are compared with the time-dependence of the nonequilibrium force averages obtained with $v_0$=0.3 in Fig.3b. Friction forces on 1 and 2 have very different transient behaviour, and do not vanish in the long-time limit, but converge to values of equal magnitude and opposite direction.

Fig.3c shows nonequilibrium forces against the liquid velocity. In the steady state, the velocity of the liquid centre of mass is $v_0/2$ in the direction of motion of the moving particle. The magnitude of relative velocity of the two particles with respect to the liquid is the same in the steady state, but the directions are opposite, resulting in opposite forces of friction.

When particle 1 starts moving, it instantly has the full ($v_0$) velocity difference to the surrounding liquid and experiences the full friction force. The stationary particle 1 is at the same time at local equilibrium with the liquid around it. As the disturbance spreads from the region around 1 to the region around 2, the friction force on 2 increases – this is why the friction force on 1 decays, and the friction force on 2 increases, to the steady state value.

It should be mentioned that the friction force on the two particles in principle depends on their separations and on the direction of their relative motion with respect to their displacement vector. (In equilibrium, this would depend on the choice of direction of the constant velocity $\mathbf{v}_0$). For sufficiently small separations, this would be reflected in the periodic time dependence of the friction coefficients in nonequilibrium simulation results as the moving particle crosses the periodic cell. If the final state showed such time dependence, one would need to use the formalism described in [12] and [13] in order to reproduce the real time-dependence of friction coefficient from equilibrium fluctuations.

## C. Particles moving with the same speed in opposite directions

If the two infinite-mass particles are moving with the same speed $v_0/2$ in opposite directions (Fig.4a), the nonequilibrium equations of motion starting at $t$=0 are

for particles 1,2: for liquid atoms ($i$=3,…,$N$):

$$\dot{\mathbf{r}}_{1,2} = \mathbf{p}_{1,2}/M \pm \mathbf{e}_x v_0/2 \qquad \dot{\mathbf{r}}_i = \mathbf{p}_i/m$$
$$\dot{\mathbf{p}}_{1,2} = \mathbf{F}_{1,2} \qquad \dot{\mathbf{p}}_i = \mathbf{F}_i.$$

**Eq. 21**

This motion creates the dissipative flux

$$J = -\sum_{i=2}^{N} (F_{x1} - F_{x2})/2V,$$

**Eq. 22**

so that the friction forces on the moving particles are

$$\frac{\langle F_{x1,2}(t)\rangle}{v_0} = -\frac{1}{2k_B T}\int_0^t \langle F_{x1,2}(s)[F_{x1}(0) - F_{x2}(0)]\rangle ds.$$

**Eq. 23**

The friction coefficients, shown in Fig.4b, converge quickly to the opposite steady-state values. The liquid centre-of-mass velocity is zero at all times, which is the reason for fast convergence, since there is no need to initially perform work in order to move the centre of mass of the liquid. The steady state values are the same as in Fig.3b,c, since the relative velocities with respect to the liquid are the same.

## D. Relationships between friction coefficients

For two infinite-mass particles immersed in a liquid, one can define four different equilibrium friction coefficients $\zeta_{a,b}$, (called "self and mutual" in Hansen),

$$\zeta_{ab}(t) = \frac{1}{k_B T}\int_0^t \langle F_a(s) F_b(0)\rangle dt.$$

**Eq. 24**

where $a$ and $b$ can each take the values of 1 or 2, and $F_{a,b}$ are the components of the forces in the chosen direction of motion.

Symmetries between the coefficients $t\to\infty$ can be easily interpreted in the context of the nonequilibrium flows in which they arise. For example, from the equality of friction forces in the long-time limit when both particles are moving together (Fig.2a), we find that for $t\to\infty$

$$\varsigma_{11} + \varsigma_{12} = \varsigma_{22} + \varsigma_{21},$$

**Eq. 25**
and because they vanish,

$$\varsigma_{11} = -\varsigma_{12}$$
$$\varsigma_{22} = -\varsigma_{21}.$$

**Eq. 26**
From the experiment in which the particles are moving in opposite directions at the same speed, where $\langle F_1\rangle = -\langle F_2\rangle$, in the long-time limit we have

$$\varsigma_{11} - \varsigma_{12} = \varsigma_{22} - \varsigma_{21}.$$

**Eq. 27**
From Eq. 25 and Eq. 27 it follows that for $t\to\infty$ self-coefficients are equal ($\zeta_{11}=\zeta_{22}$), and that mutual coefficients are equal ($\zeta_{12}=\zeta_{21}$). One could reach the same conclusion by observing the equivalence of moving particle 1 while 2 is stationary and *vice versa* (Fig.3a). These relationships were observed in [8], but their physical significance was not realized.

## E. Friction force in general motion of two particles

What would be the friction forces experienced by two massive particles at a given separation, which started moving at $t$=0 through the liquid at constant velocities $\mathbf{v}_1$ and $\mathbf{v}_2$ respectively, in terms of the linear response friction coefficients $\zeta_{ij}$ Eq. 24? (In periodic boundary conditions, the particles would mean periodic lattices.) For general relative motion, each friction coefficient would have three Cartesian components *x*,*y*,*z*, generally different from each other. We shall limit the discussion to one Cartesian direction (e.g. *x*), with components of particle velocities in this direction equal to $v_1$ and $v_2$. The motion of the two particles is a superposition of two components. One is the motion of both particles together at the velocity $(v_1+v_2)/2$ (as in Fig.2a), and the other is the motion of particle 1 at $(v_1-v_2)/2$ and the motion of particle 2 at $-(v_1-v_2)/2$ (Fig.4a). The ensemble-averaged forces in this Cartesian direction would be

$$\begin{aligned}\langle F_1(t)\rangle &= -\frac{v_1+v_2}{2}[\varsigma_{11}(t)+\varsigma_{12}(t)]-\frac{v_1-v_2}{2}[\varsigma_{11}(t)-\varsigma_{12}(t)]\\ \langle F_2(t)\rangle &= -\frac{v_1+v_2}{2}[\varsigma_{22}(t)+\varsigma_{21}(t)]+\frac{v_1-v_2}{2}[\varsigma_{22}(t)-\varsigma_{21}(t)]\end{aligned}$$

**Eq. 28**

The first term on the right hand side of Eq. 28 represents the transient effects associated with the motion of both particles at the same velocity, and vanishes at long times in both equations. Long-term friction is represented by the second term corresponding to relative motion of the two infinite masses. Eq. 1 can be rewritten in a more familiar form [8],

$$\begin{aligned}\langle F_1(t)\rangle &= -v_1\varsigma_{11}(t)-v_2\varsigma_{12}(t)\\ \langle F_2(t)\rangle &= -v_1\varsigma_{21}(t)]-v_2\varsigma_{22}(t).\end{aligned}$$

**Eq. 29**

## 4. CONCLUSION

It is well known that the correlation integral, representing friction coefficient of a massive particle moving in a liquid, vanishes in finite volume or in periodic boundary conditions. For a long time, this result has been assumed to be an unphysical effect of finite volume, not true in reality [8]. I have shown here that this is a result that reflects the real outcome of moving an infinite mass through a liquid. In this case, the infinite mass eventually drags the whole finite volume, containing a finite total mass of liquid. Thus, the reason for vanishing of the Green-Kubo expression Eq. 1 is not the finite *volume* of the liquid, but the finite *mass* of the rest of the system. The misconception that the vanishing of friction coefficient in a finite volume is unphysical has probably arisen because it was not kept in mind that in reality there are infinite mass walls bounding a finite volume liquid, preventing it from being dragged with the infinite moving mass.

Bocquet *et al* [8] studied different combinations of equilibrium friction coefficients in a system consisting of two infinite-mass particles and the rest of the liquid. I showed why some combinations vanish in the long-time limit, while some of them do not, by analysing the nonequilibrium flows of which they are the linear limit. The analysis revealed that the nonvanishing friction coefficients represent averaged forces on each particle divided by the relative velocity of the two moving particles. Friction in a liquid can exist only when there is motion of a mass relative to the rest of liquid. Such motion can exist only when at least two infinite masses are moving relative to each other.

As a corollary, the friction coefficients Eq. 24 would not vanish for a finite liquid volume between planar walls, when the immobile walls bounding the liquid represent the two infinite-mass particles [14]. Similarly, when one “particle” is a planar wall bounding the liquid, and the other is a massive particle moving past the wall, the same expressions would represent various non-vanishing friction coefficients (with the same long-time limit) that would arise in such motion. However, it is not possible to obtain the long term friction coefficient of a particle moving at a constant velocity through liquid of *finite total mass* (without immobile bounding walls) from Eq. 1.

## *FIGURE CAPTIONS*

Fig.1 (a) One infinite mass particle (black circle) moving at velocity $v_0$ in a liquid in periodic boundary conditions. Liquid (consisting of molecules in the simulation) is depicted in grey as a whole. The square represents the periodic box. (b) Comparison of the time-dependence of the ratio of the force on the moving particle and its constant velocity evaluated in equilibrium (Eq. 9 – full line) and in nonequilibrium simulations with $v_0$=0.3 (dashed line). (c) Nonequilibrium results: time dependence of the friction coefficient (full line) and average velocity of the liquid (dashed line). (d) Liquid initially (at $t$=0) moving at the average velocity $v_0$ past the immobile (infinite mass) particle. (e) Friction coefficient from equilibrium (Eq. 13 – full line) and nonequilibrium simulations with $v_0$=0.3 (dashed line). (f) Nonequilibrium results: time dependence of the friction coefficient (full line), average velocity of the liquid (dashed line) and average momentum of the liquid molecules.

Fig.2 (a) Two infinite mass particles (black and white circles) moving at the same velocity $v_0$ in a liquid in periodic boundary conditions. Liquid is depicted in grey as a whole. The square represents the periodic box. (b) Force divided by velocity on each of the particles from equilibrium simulations (Eq. 17).

Fig.3 (a) Two infinite mass particles (black and white circles) in a liquid in periodic boundary conditions. Liquid is depicted in grey as a whole. The square represents the periodic box. Particle 1 (black circles) is moving at the velocity $v_0$, while particle 2 (white circles) is stationary. (b) Force divided by velocity on each of the particles from equilibrium (Eq. 20 – full line for particle 1 and dashed line for particle 2) and nonequilibrium simulations (thin full line for particle 1 and dotted line for particle 2).

Fig.4 (a) Two infinite mass particles (black and white circles) moving at opposite velocities $\pm v_0/2$ in a liquid in periodic boundary conditions. Liquid is depicted in grey as a whole. The square represents the periodic box. (b) Force divided by velocity on each of the particles from equilibrium simulations (Eq. 23).

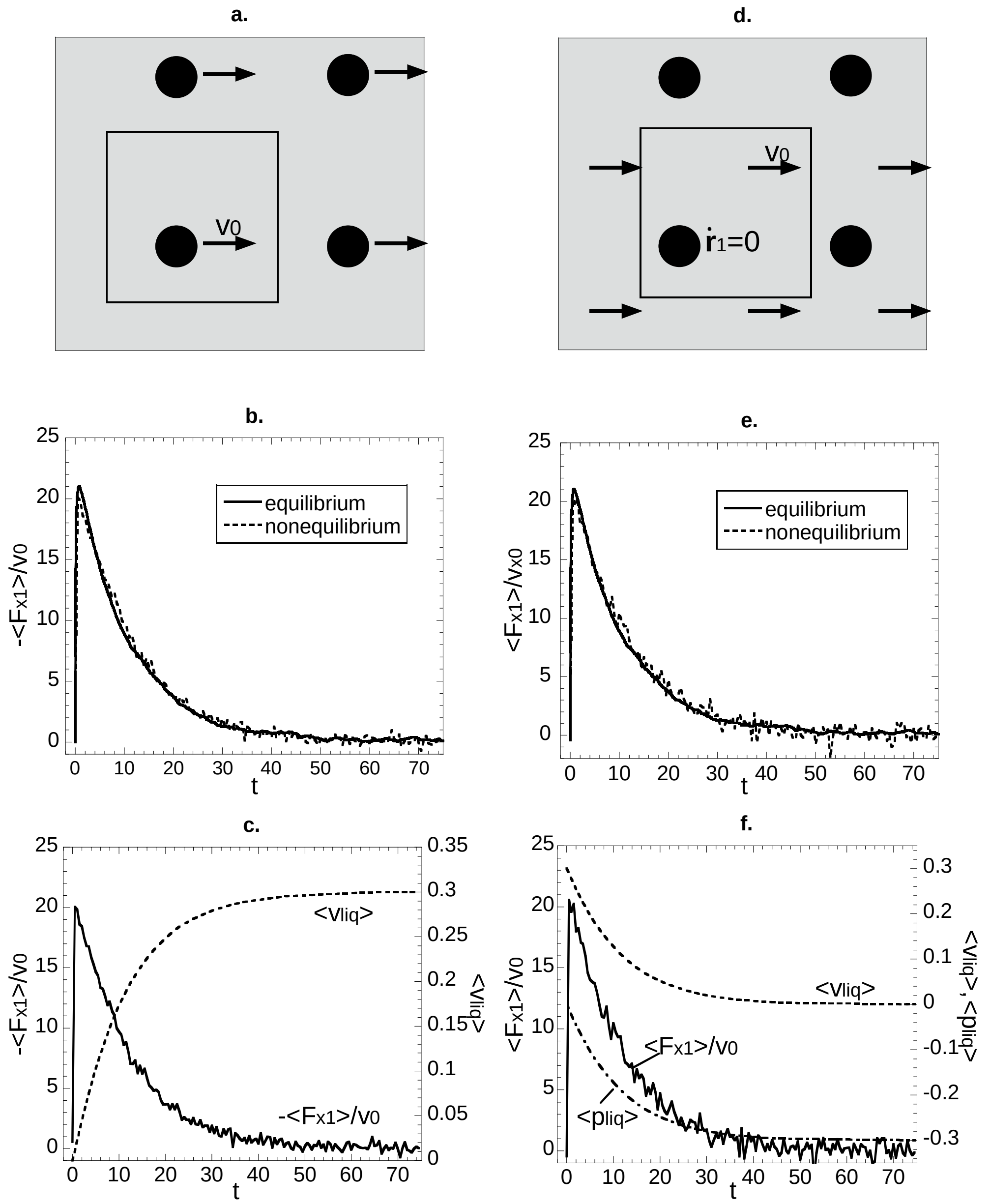

Figure 1 Petravic

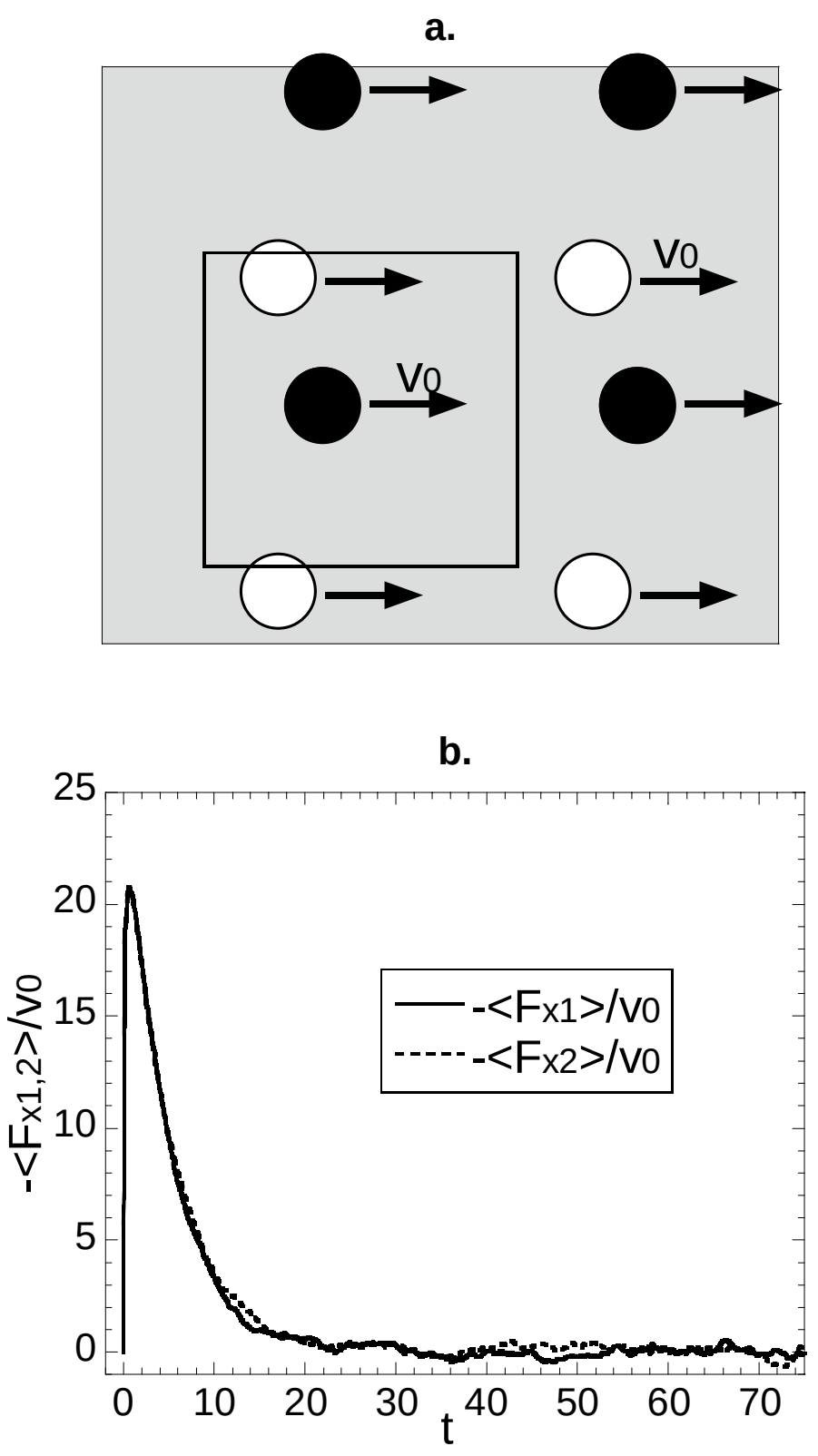

Figure 2 Petravic

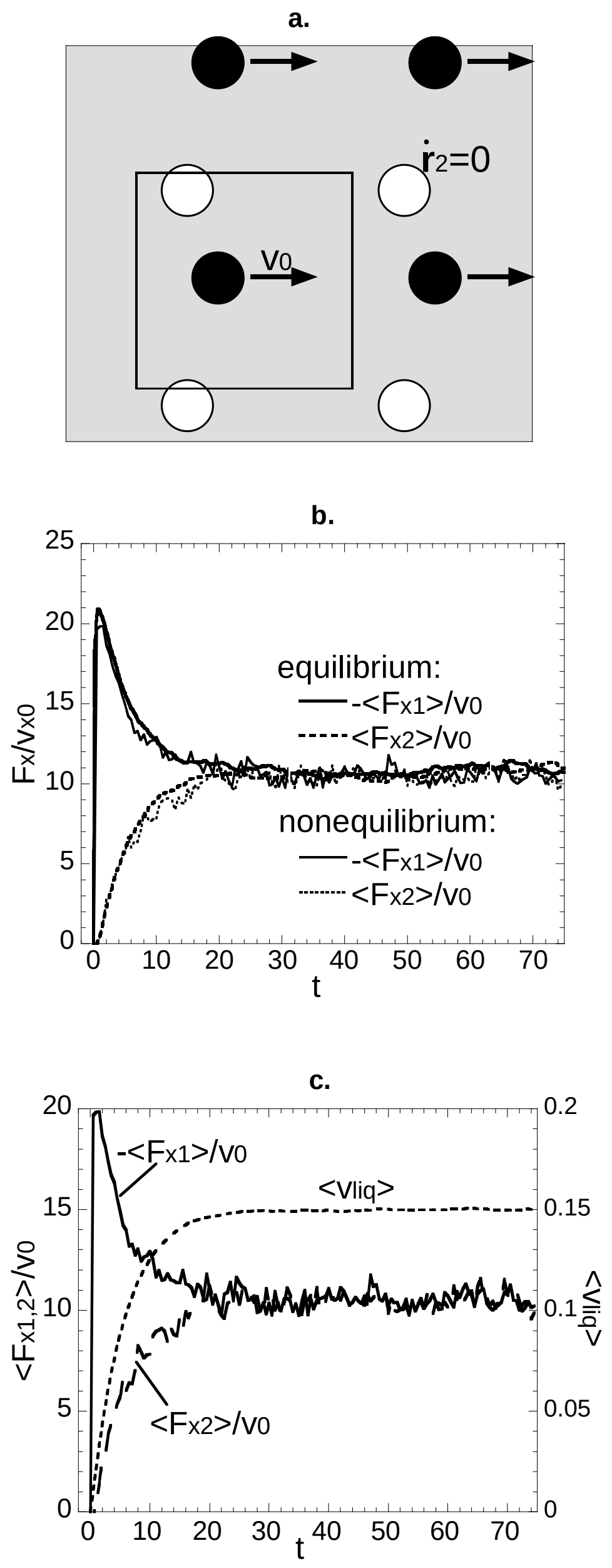

Figure 3 Petravic

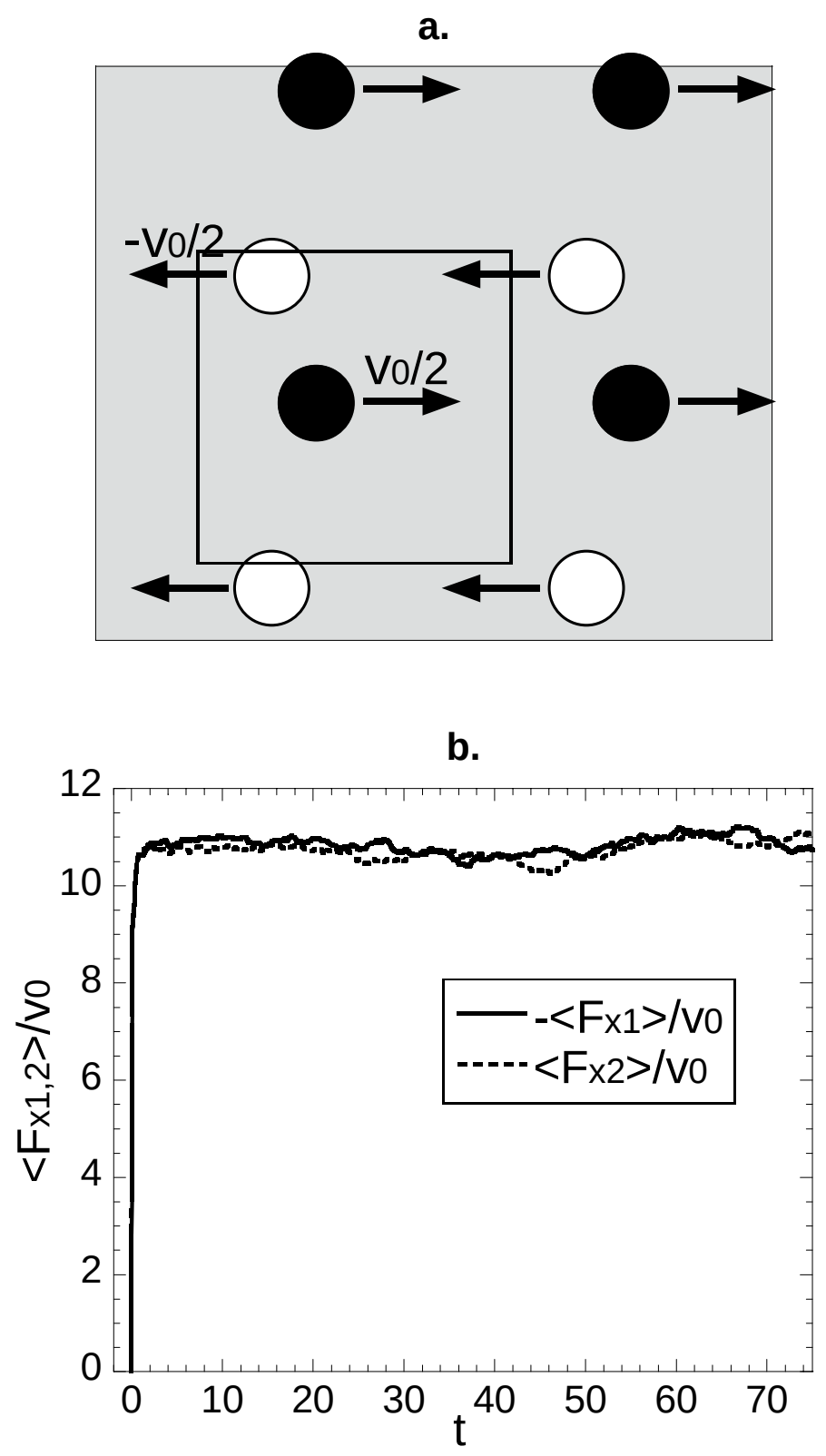

Figure 4 Petravic